%% file: dbg.tex
\documentstyle[mathptm,pramana,epsf,floats,pstricks]{ias}

\def\be{\begin{equation}}
\def\ee{\end{equation}}
\def\lsim{\raise0.3ex\hbox{$<$\kern-0.75em\raise-1.1ex\hbox{$\sim$}}}
\def\gsim{\raise0.3ex\hbox{$>$\kern-0.75em\raise-1.1ex\hbox{$\sim$}}}

\def\D#1{\displaystyle #1}
\def\PRL#1{{\sl Phys. Rev. Lett.} {\bf #1}}
\def\PRD#1{{\sl Phys. Rev.} {\bf D#1}}
\def\NPB#1{{\sl Nuc. Phys.} {\bf B#1}}
\def\PLB#1{{\sl Phys. Lett.} {\bf B#1}}

\def\vev#1{\langle #1 \rangle}
\def\SD{{\cal D}}
\def\nb{$n_{B}$}

\def\ng{$n_{\gamma }$}
\def\B{{\sl B}}
\def\C{{\sl C}}

\def\CP{$CP$}

\def\mh{$m_H$}

\begin{document}
\mark{{Baryogenesis}{U. A. Yajnik}}
\title{\Large  Baryogenesis}

\author{\large U A YAJNIK}
\address{Physics Department, Indian Institute of Technology,
Bombay, Mumbai 400\thinspace076\\
E-mail : yajnik@phy.iitb.ernet.in }

\abstract {Developments in understanding of Baryogenesis are 
reviewed. We start with early motivations and the
proposals in the context of GUTs. Next, the importance 
of the sphaleron solution and its implications are discussed.
Studies of the Standard Model reveal that the latter has a
Higgs structure incompatible with existence of observed
\B\ asymmetry. We then discuss a generic scenario for electroweak 
baryogenesis relying on bubble wall dynamics.
We also summarise the status of the MSSM,  and alternative scenarios
utilising topological defects as the source of non-equilibrium
behaviour and leptogenesis. }

\maketitle

\section{
Why Particularly, the sky?\\
{\it or} \\
Cosmology as the laboratory for Particle Physics}\label{sec:intro}

Gauge symmetry of Particle interactions and spontaneous
breakdown of the same in Unified theories
imply interesting collective phenomena at high 
temperatures. Thus in the early Universe, we expect 
phase transitions, exotic states of matter, topological
defects and so on. Some of these phenomena are expected
to leave behind observable imprints. For example Inflation
results from the unusual equation of state obeyed by vacuum
energy, or stable cosmic strings can bias the lumping of
matter. This talk will focus on baryon asymmetry resulting
from nonequilibrium conditions existing in the expanding 
Universe.

The baryon asymmetry can also be formulated as the 
baryon number to entropy density ratio of the Universe,
$\Omega=n_B/s$. From direct observation of the number of 
galaxies, the average number of stars per galaxy and so on,
$\Omega$ has the value $10^{-10}$.
This is corroborated by the relative
abundance of light nuclei in Big Bang nucleosynthesis,
$1.5\times10^{-10}<\Omega<7\times10^{-10}$.
This unnaturally small value demands a microscopic explanation. 
Intensive developments over
the past decade have yielded important information
about Particle phenomenology, mostly in the form of
vetoes. We present here a review for the non-expert.

Investigating the consistency of a given model of
Particle Physics with the observed baryon asymmetry
requires checking for the nature of high temperature
phase transition in the theory and also for existence
of requisite particle species content. We focus here
on the general nature of these requirements, how they
arise and then summarise the results for some of the
popular models. The interesting result is that most
of the models accommodate the observed baryon asymmetry
only with fine tuning of parameters.

\subsection{Asymmetry observed}
The need to understand Baryon asymmetry arises in the
first instant from the absence of any antimatter. How do
we know the asymmetry observed in the immediate astral 
neighborhood is Universal? The three broad classes of 
data in this connection are
\begin{itemize}
\item{} The content of cosmic rays is baryonic. The 
observed ratio of $\bar p$ to $p$ in cosmic rays is
$10^{-4}$. This is consistent with secondary production
$p+p\rightarrow 3p+{\bar p}$.
\item{} If neighboring clusters of galaxies happened to 
contain matter and anti-matter, this would produce
diffuse $\gamma$-ray background. This is not observed.
\item{} Perhaps the regions of anti-matter are completely
separated. This would reflect in inhomogeneities in the
Cosmic Microwave Background radiation (CMBR). These 
are also not observed.
\end{itemize}

We begin in section \ref{sec:orig} with a review of the 
general requirements for baryogenesis (\B-genesis) followed
by the more specific receipe of Grand Unified Theory (GUT) 
baryogenesis. Section \ref{sec:sphal} contains the
significance of the sphaleron solution and the beginning
of the modern attack on the problem. It is shown that
the Standard Model (SM) Higgs structure is incompatible with
observed baryon asymmetry on fairly general grounds.
Section \ref{sec:ewbgen}
discusses the potential for  \B-genesis at the electroweak 
scale in extensions of the SM,  with the example of the two 
Higgs doublet model (2HDM).
The current status of the MSSM is also discussed.
In section \ref{sec:topalter} we present briefly 
the status of other mechanisms, viz., \B-genesis induced
by topological defects, and \B-genesis through 
Leptogenesis\cite{fukyan}.
Section \ref{sec:conclusion} contains the conclusion.

\section{Baryogenesis in the beginning}\label{sec:orig}

Several peculiar features of the nuclear interactions 
such as Parity and \CP\  violation became known by the early
1960's. The discovery of the Cosmic Microwave Background
(CMBR) around the same time was confirming a cosmological 
arrow of time. Another puzzle that was noted around this 
time was the fact that Baryon number, a symmetry
of the strong and weak forces was only an algebraic
symmetry and not a gauge symmetry which would have a fundamental 
justification for its exact conservation. 
Combined with \CP\  violating interactions
and the the expanding Universe this presented the possibility
of an {\it explanation} from physical laws of the Baryon 
asymmetry\cite{wein1}.  A first explicit model by Sakharov\cite{sakh}
set forth the following salient issues, the so called
Sakharov criteria : 1. \B\  violating and 2. \C\  violating interactions,
3. \CP\  violation 4. Out of equilibrium conditions, i.e.,
the state of the system must be time asymmetric.
The last ensures that \CP\  violation becomes effective. 
We recapitulate here the model of Yoshimura\cite{yosh} 
and Weinberg\cite{wein2} proposed in the context 
of the GUTs.

\begin{description}

\item[Baryon number violation]\ 
 Consider a species $X$
that has two different modes of decay (even if the absolute
baryon number $X$ is not defined, it is the difference in the
baryon number of the final states that matters).

\begin{eqnarray*}
X \quad &\longrightarrow& \quad qq \qquad \Delta B_1 = 2/3\\
&\longrightarrow& \quad {\bar q}{\bar l} \qquad \Delta B_2 = -1/3
\end{eqnarray*}

\item[Charge conjugation violation]\ 
The inequality of the amplitudes for the charge conjugated processes,
${\cal M}(X\rightarrow qq)\neq{\cal M}({\bar X}\rightarrow 
{\bar q}{\bar q})$

\item[$CP$\  violation]\
\[
r_1={\Gamma_1(X\rightarrow qq)\over\Gamma_1+\Gamma_2}
\neq{{\bar \Gamma}_1({\bar X}\rightarrow {\bar q}{\bar q})
\over{\bar \Gamma}_1+{\bar \Gamma}_2}={\bar r_1}
\]
It is clear that phases appearing in the vertices of an effective 
Hamiltonian cannot enter in the rate formulae in the Born approximation. 
A crucial observation of \cite{wein2} was that it must be the 
interference of a tree diagram with a higher order diagram with 
a \CP\ violating phase which will result in $r\neq{\bar r}$.

\item[Out of equilibrium conditions]\ 
The effect of the dominant forward reaction would be
nullified by excess build-up of the species which would
drive the reverse reaction and
establish an equilibrium with the anti-species unless
the state of the system prevented this from happening.
\end{description}

If these conditions are satisfied the rate
of Baryon number violation is 
\begin{eqnarray}
{\cal B} &=& \Delta B_1 r_1 + \Delta B_2 (1-r_1) + (-\Delta B_1){\bar r_1}
 + (-\Delta B_2)(1-{\bar r_1}) \nonumber \\
 &=&\ (\Delta B_1 - \Delta B_2)(r_1-{\bar r_1})
\label{eq:bformula}
\end{eqnarray}

The decay rate for the $X$ particle in the early Universe 
has the temperature dependence $\propto 1/T$ while the 
Hubble parameter $H$ varies as $\propto T^2$. The condition 4
above comes to be satisfied when $\Gamma \lsim H$,  
which happens in GUTs very early in the Universe.
The resulting $\Omega$ is estimated \cite{wein2}  to be  
$\sim {\cal B}/g_*$ where $g_*$ is the effective number
of degrees of freedom in equilibrium at that epoch.
In order to obtain quantitative results 
the Boltzmann equations must be integrated taking
account of each decay mode and the chemical potential
of each of the conserved numbers.

This GUT based scenario has several problems if inflation has
to also occur. However, more recently, the same scenario 
has been applied to the lepton number violation by
heavy majorana  neutrinos \cite{fukyan}, as discussed in 
sec. \ref{sec:topalter}.
In the meantime, an important discovery of the mid-
1980's has completely revolutionized our understanding
of the fate of the Baryon number at a much lower energy
scale, viz., the electroweak scale. This is what we turn 
to next.

\section{The anomalous baryon number}\label{sec:sphal}

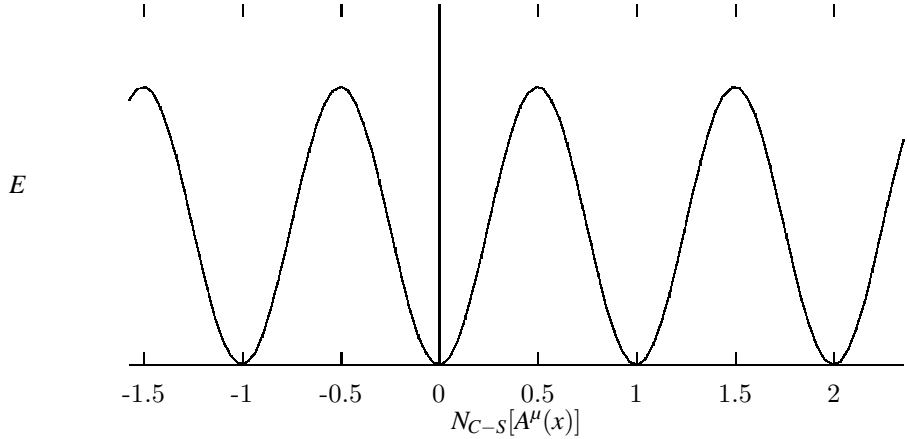
\begin{figure}
\input{vacgge.tex}
\caption{Energy profile of gauge field configurations}
\label{fig:eprofile}
\end{figure}

Since the Standard Model is chiral it has the potential
presence of anomalies.
All the gauge currents and most of the global conserved
currents are indeed anomaly free in the Standard Model. 
However, the number $B+L$ is in fact anomalous.
Specifically, one finds that

\begin{equation}
\partial _{\mu }j^{\mu }_{B+L} = {g^{2}\over 32\pi ^{2}} 
\epsilon ^{\mu \nu \rho
\sigma }F_{\mu \nu }^aF_{\rho \sigma }^a
\label{eq:anomaly}
\end{equation}

\noindent where $g$ is the gauge coupling. This anomaly of 
the  fermionic current
is associated with another interesting fact of non-Abelian 
gauge field theory discovered by Jackiw and Rebbi.
There exist configurations of the gauge
potentials $A_\mu^a$ each of which is pure gauge, i.e., 
with physical field strengths $F_{\mu \nu }^a=0$, but which cannot 
be deformed into each other without turning on the physical
fields. Such  pure  gauge  vacuum   configurations   can   be
distinguished from each  other  by  a  topological  charge  
called  the Chern-Simons number
\par
\be
N_{C-S} ={g^3\over 32\pi^2} \epsilon ^{ijk} \int  d^{3}x 
\left\{F_{ij}^aA^a_k
 - {2\over 3}\epsilon ^{abc}A^{a}_i A^{b}_j A^{c}_k
\right\}
\label{eq:NCS}
\ee
which has integer values in vacuum configurations.
On the other hand the RHS of the anomaly eqn. (\ref{eq:anomaly}) 
is equal to a total divergence $\partial _{\mu}K^{\mu }$
where
\par
\be
K^{\mu } = \epsilon ^{\mu \nu \rho \sigma } \left\{
F^{a}_{\nu
\rho } A^{a}_{\sigma }\,-\,{2\over 3}\,g\,\epsilon
^{abc}A^{a}_{\nu } A^{b}_{\rho }A^{c}_{\sigma }\right\}
\label{eq:kmu}
\ee
\noindent Hence if we begin and end with configurations with
$F^{a}_{\mu \nu}=0$, then the change
\be
\Delta Q_{B+L} \equiv  \Delta  \int  j^{0}_{B+L} d^{3}x
= - \Delta \left( {g^{2}\over 32\pi
^{2}} \int K^{0} d^{3}x \right)
= - \Delta N_{C-S}
\label{eq:deltaq}
\ee
\noindent
using (\ref{eq:anomaly}) (\ref{eq:NCS}) and (\ref{eq:kmu}).
Thus the violation of the axial charge by unit occurs because of a
quantum transition from one pure gauge configuration to another.
At normal temperatures such transitions do not occur. As shown
by 'tHooft, the tunneling probability is highly suppressed 
due to instanton effects. In case of gauge Higgs system with
symmtry breaking, a barrier exists between the equivalent vacua.
There is a configuration called the  sphaleron\cite{vsoni}
\cite{kliman} which is  supposed  to   be   a   time
independent configuration of gauge  and  Higgs  fields  which  has
maximum energy along a minimal path joining sectors  differing  by
unit Chern-Simons  number. In figure \ref{fig:eprofile} the
sphaleron corresponds to the peaks where $N_{C-S}$ has half-integer
values. It provides a lower bound on the amount of energy required
for passage {\it over} the barrier.
The sphaleron solution of the field equations has been discussed
in detail\cite{brihaye} in many references. Here we shall focus on 
its consequences.

\subsection{The implications of the sphaleron}

The energy of the sphaleron $E_{sph}$, can be calculated from numerical
solution of the gauge field equations satisfying the appropriate
boundary conditions \cite{kliman}. It is a weak function of
$\lambda/g^2$, i.e., of $m_H/m_W$ in the SM. It varies over
$7$ to $14$ TeV as this parameter varies from small values to
infinity. At moderately high temperatures, by which is meant
$m_W\ll T\ll m_W/{\alpha_W} \sim E_{sph}$, the barrier crossing 
rate is suppressed by a Boltzmann factor, and in saddle point
approximation, the rate per unit four-volume is\cite{armc}
\be
\Gamma = A\kappa ({\cal N}{\cal V}) T^4 e^{-E_{sph}(T)/T}
\label{eq:armcrate}
\ee
Here the ${\cal N}$ and ${\cal V}$ denote contributions
from zero-mode integration, $\kappa$ is the contribution of
fluctuations and $A$ is a dimensionless prefactor. 

Suppose we have some
mechanism for producing baryons above the electroweak scale.
Once the symmetry breakdown occurs in the early Universe,
sphalerons become possible. Then \B\  number begins to deplete
at a rate determined by the above rate (\ref{eq:armcrate}).
In order for the \B-asymmetry to survive we need that the
above rate is really too slow compared to the expansion rate 
of the Universe. We refer to this as the ``wash-out'' constraint,
and is the single most important implication of sphaleron 
physics, following entirely from the SM\cite{mhbound}. As a
function of temperature,  $E_{sph}$  is given parametrically by
\be
E_{sph}(T) \sim B {m_W(T) \over \alpha_W(T)} \propto {v(T_c)\over T_c}
\label{eq:esphT}
\ee
where $B$ is a dimensionless quantity of order $1$ and
$v(T)$ is the temperature dependent vacuum value of the Higgs. 
$E_{sph}$ should be large enough to prevent wash-out, which
requirement is shown numerically to translate to
\be
 {v(T_c)\over T_c} \gsim 1
\label{eq:fopt}
\ee
Since the order parameter has zero vacuum value before the phase 
transition, large $v(T_c)$ immediately after means that the 
phase transition should  be strongly first order. 
From perturbative effective potential one
learns that the SM phase transition is only mildly first order for
the experimentally acceptable range \mh$\gsim 90$ GeV, for the 
SM Higgs. {\it Thus the  SM  is contradicted by the presence 
of \B-asymmetry in the Universe}, unless $B-L$ is also 
not conserved, or that a net primordial value of this number
has pre-existed the electroweak phase transition. In either
case, the SM is demonstrated to be insufficient for a
physical explanation of the \B\ asymmetry.

Above arguments relying on saddle point perturbation theory 
are physically transparent but only suggestive.
More recently, a non-perturbative approach using 
lattice methods has been developed\cite{moorelat} 
and so far seems to not contradict the above conclusions.

\subsection{The anomaly at higher temperatures}
It is also natural to ask about the physics of the anomaly
above the electroweak temperatures. With \mh $\sim 120$ to 
$150$ GeV, the phase transition temperature $T_c$ is $100$
GeV. Above this temperature, $v(T_c)$ vanishes. Then there
is no sphaleron and it is reasonable to assume that there
is no barrier either. Assuming that coherent fluctuations
can exist on the scale of the magnetic screening length
in the non-Abelian plasma, the rate of $N_{C-S}$ changing
transitions is given in dimensional analysis by
\be
\Gamma = C (\alpha_W T)^4
\ee
In fact there are recent arguments to the effect that the prefactor
$C$ is $\sim \alpha_W$. In either case, there is no suppression
and we assume that $B+L$ is freely violated. This conclusion
on general grounds has been verified by real time simulations 
in lattice gauge theory \cite{aaps}. It is shown 
that the $N_{C-S}$ value oscillates around a given integer value
corresponding to a specific potential well, but every once
in a while, makes a sharp transition to a neighbouring well.
The rate of transitions is then found to accord with above formula.

This raises an important issue in the explanation of baryon
asymmetry of the Universe. $B+L$ asymmetry generated by any 
mechanism above the electroweak scale will be wiped out
by the unsuppressed anomalous transitions at high temperatures.
There are two solutions to this dilemma. One is to find 
physical mechanisms that produce a net $B-L$. A very attractive
candidate in this class is the \B-genesis through Leptogenesis
scenario to be discussed in section \ref{sec:topalter}.
A tantalising possibility first suggested by Kuzmin, Rubakov
and Shaposhnikov\cite{krs} is the possibility of processes
operating at the
electroweak scale itself that would produce the required
baryon asymmetry. Since the interactions involved are or will
soon be within the range of accelerator energies, this 
possibility is very exciting. Appropriately, it generated
a decade long intense search for mechanisms that would work
at electroweak phase transition temperatures, as discussed
in the following section.

\section{Electroweak baryogenesis}\label{sec:ewbgen}

Any mechanism for \B-genesis 
operating at the electroweak temperatures differs in an important
way from the Yoshimura-Weinberg proposal in that we need
a new source of time asymmetry. The point is that
at the GUT scale the expansion of the Universe is 
fast enough to exceed the decay rates of relevant 
particle species, while at the electroweak epoch the
expansion rate is orders of magnitude smaller.
Any decays which are out of equilibrium at this 
epoch will in any case prove insufficient to the task.

\begin{figure}
\epsfxsize=4cm
\centerline{\epsfbox{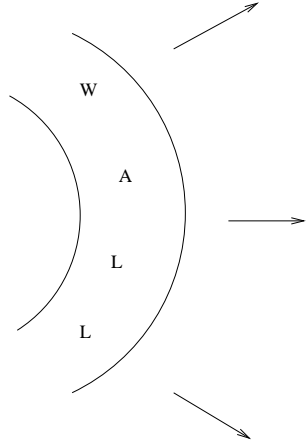}}
\vskip 3mm
\caption{Cartoon of a bubble wall. The region
to the left has $\vev{\phi}\neq0$ while the region to the 
right into which the bubble is expanding has $\vev{\phi}=0$.
The region labelled WALL has interpolating values of 
$\vev{\phi}$}
\label{fig:bubexpand}
\end{figure}

A novel feature of the the electroweak physics is that
the symmetry breaking transition could be
first order, in the sense that the transition involves 
latent heat and the true vacuum is formed by tunneling
from the false vacuum. Such formation would typically
proceed in the shape of spontaneously forming ``bubbles''
as per the mechanisms of Coleman \cite{colbub} and
 Linde\cite{linbub}. 
See figure \ref{fig:bubexpand}. In the 
context of this picture we expect
the expanding bubble walls to provide the out of equilibrium
conditions needed for baryogenesis.

We now discuss this in greater detail.
What needs to be verified is whether the given phase transition
is indeed first order. One way to study this is to study
the effective potential at finite temperature. The free energy of the
Higgs field is given in the field theoretic formalism by the
finite temperature effective potential. The one loop contribution
to the effective potential of the Higgs from integrating out 
other particles is of the form 
\be
\Delta V_T^i \sim T^4 \int dx x^2 \ln (1 \pm e^{-\sqrt{x^2+y_i^2}})
\ee
where $-$ is for bosons and $+$ for fermions, and for 
each species $i$, 
\be
y_i=M_i\vev{\phi}/v_0T,
\label{eq:exppar}
\ee

$M_i$ being the respective mass and $v_0$ is
the zero temperature expectation value of the Higgs field,
$v_0=246$GeV. If $y_i$ can be treated as small parameters,
the effective potential for the Higgs becomes
\be
V_{eff}^T[\phi]\ =\  D(T^2-T_0^2)\phi^2 - ET\phi^3 
+ {\lambda_T\over 4}\phi^4
\ee
where $D$, $T_0$ and $E$ (all positive) are parameters 
determined in terms
of the zero temperature masses of the gauge bosons and
the top quark. $\lambda_T$ is only mildly temperature dependent.
At $T\gg T_0$, only the $\phi=0$ minimum exists.
For $T<T_0$, this minimum is destabilized.
But there exists a $T_c>T_0$, when another minimum  
with nontrivial value of $\phi$ becomes possible, and
\be
{\phi(T_c) \over T_c}\quad=\quad {\D 2E \over \lambda}\quad \simeq 
\quad {\D m_W^2 \over m_H^2}
\label{eq:tempvev}
\ee
A barrier separates the two vacua and  tunneling across the 
barrier via thermal and quantum fluctuations becomes possible. 
Whenever
tunneling to the true vacuum occurs in any region of space, it
results in a bubble. 
According to a well developed formalism \cite{linbub}, the
tunneling probability per unit volume per unit time is given by
\be
\gamma = CT^4e^{-S_{bubble}}
\ee
where $S_{bubble}$ is value of
\be
S = 4\pi\int r^2dr\{{1\over2}\phi^{\prime 2} +
V^T_{eff}[\phi]\}
\ee
extremised over $\phi$ configurations which satisfy the
bubble boundary conditions $\phi(r=0)=\phi(T_c)$, $\phi
\rightarrow 0$ as $r\rightarrow\infty$. Once a bubble forms,
energetics dictates that it keeps expanding, converting more of
the medium to the true vacuum. The expansion is irreversible and
provides one of the requisite conditions for producing
baryon asymmetry. 

\subsection{Two obstacles to SM \B-genesis}
Thus the SM possesses in principle all the ingredients 
necessary for producing \B-asymmetry. But there are two important
issues to be faced.
First we note that the $CP$ violation available in
SM is far too small. A model independent
dimensionless parameter characterising the scale of this 
effect has the value\cite{jarlskog} $\delta_{CP}\sim 10^{-20}$.
Thus the explanation of 
\B-asymmetry is not possible purely within the SM.

But even assuming that there may be non-SM sources of $CP$ 
violation, an important question is whether the phase
transition at the electroweak scale is first order or second
order.  In the formalism discussed above we used $V^T_{eff}$
extensively. Unfortunately,
perturbative $V^T_{eff}$ is not always a reliable 
tool for studying the nature of the phase transition.
One way to understand this problem \cite{arnold} is to observe
that the expansion parameters in (\ref{eq:exppar}) are not
small. Specifically, the gauge boson contributions are
expanded in powers of (number)$\times$$m_H^2/m_W^2$. Since the
ratio of the masses is now known to be at least unity, the
validity of the perturbation series hinges crucially on
numerical factors. The only way to test the validity would be
to carry the expansion to higher orders. In these studies, 
it is found that the two-loop corrections
are not small compared to the one-loop contributions
\cite{arn2loop}\cite{buch2loop}. 

Several other approaches have been tried. One is to improve
the perturbation theory. This is done by the technique
of dimensional reduction possible in high temerature
expnasion. This method exploits the effective three dimensional
nature of the Euclidean thermal theory if only the zero
Matsubara frequency modes are retained \cite{appisar}\cite{nadkarni}.
The higher frequency modes can be accounted for in a modification
of this approach by including additional effective terms
\cite{faracetal}. The most direct approach involves 
lattice simulations, both in $4\,d$ theory and the effective
$3\,d$ theories.  The upshot of these is that the SM electroweak
pahse transition is mildly first order for \mh$\sim 100$ 
GeV, but becomes second order for highr \mh. Since the
accelerator limits on \mh\  have exceeded $90$ GeV, we clearly
need physics beyond the SM, both for the \CP\  violation as
also for obtaining out of equilibrium conditions.

\subsection{Electroweak baryogenesis in other models}
We seek an extension of the SM that satisfies the requirements
spelt out above.
In any model which suits this purpose a detailed mechanism
is necessary to bring the non-equilibrium behaviour of the
bubble walls into play. In particular, completely different
mechanisms would be successful depending upon the nature of 
the bubble walls. There are two broad classes :
\begin{description}
\item[{\sl Fast moving, thin walls}] $\qquad $ If the wall
thickness is less than the inverse thermal mass of some particle
species,  the interaction of the latter  with the wall 
can be treated in the sudden approximation. The scenario
considered by Cohen Kaplan and Nelson \cite{cokaneprop} involves 
scattering of neutrinos from the axpanding walls in this
approximation. 
Lepton number violation occurs if the neutrino obtains
a majorana mass from the scalar forming the wall. \CP\ phase
enters from the majorana coupling, and the resulting
rate of reflection for $\nu_L$ can be different 
from that for the \CP\  conjugate state $\bar \nu_R$.
Finally the excess $L$ number generated in front of the wall 
is converted to \B-number by the unsuppressed anomalous 
transitions which continue to set $B+L=0$ . Once 
the \B\  excess is engulfed by the interior of the bubble, 
it is protected if sphalerons are  sufficiently heavy.

\item[{\sl Slow moving, thick walls}] $\qquad$ 
If the wall thickness is large compared to thermal
correlation langths, coherent gauge and scalar fluctuations
are possible within the walls. Thus Chern-Simons number
changing transitions are possible within the thickness
of the wall. Further, as recognised by  McLerran,
Shaposhnikov, Turok and Voloshin\cite{mstv}, the nontrivial 
background field of the wall induces a non-trivial 
chemical potential for the $N_{C-S}$ as discussed later. 
If the scalar sector
of the theory also contains \CP\  violation then we may
expect a nontrivial asymmetry to result. The \B\  number 
thus generated falls in the interior of the bubble and is 
again assumed to not be depleted significantly by the
spahlerons of the theory.

\end{description}

\begin{figure}
\epsfxsize=7cm
\centerline{\epsfbox{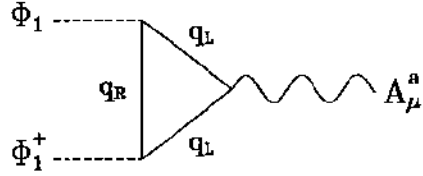}}
\caption{A nontrivial contribution to the $S_{eff}$}
\label{fig:anodia}
\end{figure}

\subsection{Baryogenesis in 2HDM}
We shall now discuss the second scenario in greater
detail. A mechanism for biasing $N_{C-S}$ is available
in scalar couplings but to achieve the desired \CP\  violation
we need the two Higgs doublet model (2HDM).
Consider the diagram of fig. \ref{fig:anodia}. 
The heavy quark in the loop is assumed to couple only to $\phi_1$
to avoid flavour changing neutral currents.
This diagram induces in the effective action a term which has 
a nontrivial expectation 
value only where the fields are space-time dependent. 

\be
\Delta S = {-7\over 4}\zeta(3) \left( {m_t\over \pi T}
\right)^2 {g\over16\pi^2}{1\over {v_1}^2}
 \times\int(\SD_i\phi_1^{\dag}\sigma^a
\SD_0 \phi_1 - \SD_0\phi_1^{\dag}\sigma^a
\SD_i\phi_1)\epsilon^{ijk}F^a_{jk}d^4x
\ee
where $m_t$ is the top quark mass, $\zeta$ is the Riemann zeta
function, and the $\sigma^a$ are the Pauli matrices.
Assuming the time derivatives dominate compared to space 
derivatives, in the gauge ${A_0}^a=0$, we can rewrite this in the
form
\begin{eqnarray}
\Delta S\ &=&\  {-i7\over4}\zeta(3)  \left({m_t\over \pi
T}\right)^2 {2\over {v_1}^2} \nonumber \\ 
& & \phantom{\Delta S\ =\ {-i7\over4}\zeta(3)} \times
\int dt [{\phi_1}^{\dag}\SD_0\phi_1 -
(\SD_0\phi_1)^{\dag}\phi_1]N_{CS} \\
&\equiv&\  {\cal O}N_{CS} \nonumber
\end{eqnarray}

This leads to\cite{mstv} an estimate of \nb/\ng$\sim 
10^{-3}\alpha_W^4 \sin 2\xi(T_c)$ where $\xi$ is the value
of \CP\  violating angle at the relevant temperature.
If $\sin 2\xi(T_c)$ is $O(1)$, this leads
to an answer in the correct range of values. 
It may be noted that the mechanism in spirit is very similar to
the generic ``spontaneous baryogenesis'' proposal of Cohen
Kaplan and Nelson\cite{cokanerev}, but which did not consider the
specifics of a phase transition bubble wall.

\begin{figure}[t]
\epsfxsize=7cm
\centerline{\epsfbox{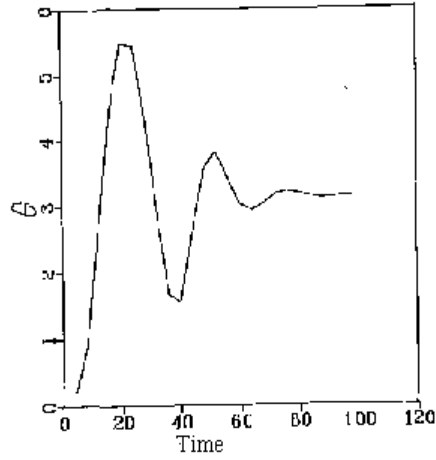}}
\caption{Oscillations of the \CP\  phase at a point swept 
by the advancing wall}
\label{fig:cpphase}
\end{figure}

In \cite{mstv} it was assumed that the relative phase 
between the vacuum expectation values of the Higgs remains
static. This however is too restrictive and leads to
a suppression of this effect in the lowest adiabatic order as 
noted by the authors. An appropriate generalisation is to let
this phase $\theta$ be time dependent and 
make the ansatze\cite{sbdanduay2}
 \begin{eqnarray}
 {\phi_{1}}^0 &=& {\rho}(r,t)\cos{\gamma} e^{-i\theta(t)}  \\
 {\phi_{2}}^0 &=& {\rho}(r,t)\sin{\gamma} e^{i\omega(t)}
 \end{eqnarray}
Here $\gamma$ is the particular direction in the ${\phi_{1}}^0 - 
{\phi_{2}}^0 $ plane along which the scalar fields tunnel
to form a bubble.
The angles $\omega$ and $\theta$ are related by the unitary
gauge requirement that the Goldstone boson eaten by the $Z^0$
should be orthogonal to the remaining physical pseudoscalar:
$\partial_\mu \omega =$$ {(\rho_1 / \rho_2)}^2 \partial_\mu \theta$
\cite{cokanerev}. The variation of $\theta$  
as the wall sweeps past a particular point is shown in 
fig. \ref{fig:cpphase}\cite{sbdanduay2}. The first peak of this graph occurs
over the time scale it takes for the wall to sweep past 
a particular point. The remaining oscillations occur in the
wake of the wall. Thus, ${\rm Im}\phi$ which was static and gave 
no leading order contribution is now replaced by a phase
which oscillates over its full range of values. This $\Delta\theta$
acts as the transient \CP\  violation parameter.

To estimate the generated baryon asymmetry we proceed as in 
the general case outlined earlier. The number of fermions
created per unit time in the bubble wall is given by
     \begin{equation}
     B = \kappa {(\alpha _{w} T)}^{4} l S\times
{\frac {1}{T}}{\frac {{\cal O}} {l}}
     \end{equation}
where\cite{mstv} \cite{cokanerev}, $l$ and $S$ are the 
thickness and the surface area of the bubble
wall respectively. From which we get the baryon to photon ratio to be
\be
 \bigtriangleup\  \equiv\  {\frac {n_{B}}{s}}\  \simeq\  {\frac{1}{g_*}}
     {(\alpha_W)}^4 {\cal O } \ 
     \simeq\  10^{-8}\times \left( \frac{E}{K_1} \right)^2 \Delta\theta
\ee
where we have used $\alpha_W$ $\sim$ $10^{-\frac{3}{2}}$ and $g_* \sim
100$. The parameters $E$ and $K_1$ are given in terms, respectively,
of the cubic and quartic couplings in the 2HDM.
This answer easily accommodates the observed value of this number.

It is worth emphasizing the physics of this answer which
is robust against changes in the specific particle physics model
\cite{sbdanduay3}. 
The thermal rate contributes $10^{-6}$ through $\alpha_W^4$, 
and another $10^{-2}$ is contributed by $N_{eff}$. 
Further, the temperature
induced cubic coupling $E$ is generically small compared to $K_1$
so that in  ``bubble wall'' scenarios we would never need \CP\  
violation larger than $10^{-3}$, and more likely much less.

\subsection{Baryogenesis in the MSSM}

The Minimal Supersymmetric Standard Model is an attractive 
extension of the SM. Extensive investigations\cite{cqriw},
\cite{cqw}, \cite{clijoka} have been made
of this model leading to  reasonably definitive 
conclusions about the viability of \B-genesis in it. The sphaleron
does occur\cite{mooaqu}
in the model and high temperature anomalous processes are also
expected to occur. As was seen in the SM case, if the Higgs
is too heavy it tends to make the phase transition second
order. In MSSM, the phase transition is first order if the 
lightest Higgs and the scalar superpartner of the top quark,
the stop are sufficiently light. The \CP\ violation cannot 
arise in the Higgs sector but can occur in the soft 
SUSY breaking parameters associated with mixing of 
Left and Right stops. However, too large a mixing
between L and R stops reduces the strength of the 
first order phase transition. The baryon creation mechanism 
relies on wall motion and the interaction of \CP\  odd 
currents with the wall. It is shown that the 
generated \B\  asymmetry is proportional to change in the
$\tan\beta$ parameter (the ratio of the vacuum values of 
the two Higgs) across the bubble walls. If this variation
has to be significant, say $\sim10^{-2}$ then $m_A$, the 
mass of the pseudoscalar Higgs $\lsim 150 -\ 200$ GeV.
On the other hand, a value of $m_A$ less than above range
tends to weaken the strength of the first order phase
transition.  Thus baryogenesis is viable in the model in
a fairly restricted window in the parameter space.
Numerical calculations then show that the
lightest Higgs should be in the range $85\ -\ 110$ GeV,
accesible to LEP2 and the Right stop mass in the range 
$120\ -\ 172$ GeV. Further, $\tan\beta$ should be
$\lsim 4$.  Since the perturbative consistency of the model
upto GUT scales requires $\tan\beta\gsim 1.2$, this is a 
significant constraint. 

\section{Other Approaches}\label{sec:topalter}
\begin{description}
\item{\sl Topological defects}

It was pointed out in \cite{sbdanduay1} that the presence
of cosmic strings at the electroweak phase transition would
simplify some of the nettlesome issues related to bubble
wall thickness and velocity. The bubbles described by
the Coleman-Linde formalism are distributed in their sizes
and the times of their occurance. The tunneling probability
is a very sensitive function of the parameters in the potential
and it is not possible to work with a generic \B-genesis
scenario but each model must be examined in detail. By
contrast, if the phase transition is induced by the presence
of cosmic strings then the parameters of the bubbles are
{\sl decided entirely by electroweak physics} regardless
of the detail of the unified model giving rise to the
strings. 

A more direct role for cosmic strings was proposed in\cite{branden1}
where unstable cosmic strings themselves act as locations of 
non-equilibrium processes. A set of model independent
defect induced scenarios was also proposed\cite{troddenetal}, 
\cite{branden96}.
Recent numerical work\cite{mooreclineetal}  
shows that of these the string mediated mechanism is not efficient 
enough at \B-asymmetry production. However, the possibility that 
unstable domain walls could lead to \B-genesis is still
open \cite{mencoo} \cite{lewrio}. 
In particular, many interesting topological objects 
including unstable domain walls have been shown to occur
in the Left-Right symmetric model \cite{ywmmc} where
such scenarios can work. 

\item{\sl Leptogenesis}

The early proposal of Fukugita and Yanagida of \B-genesis from 
Leptogenesis gains a tantalising prospect after the recent 
strong indications that the neutrino has mass. Models that 
accommodate the indicated minuscule masses should incorporate
the see-saw mechanism. If this is true there must be heavy
neutrinos possessing majorana masses. Further the Yukawa
couplings giving rise to such masses can themselves be
\CP\  violating. We then have all the ingredients necessary
to play out the scenario of \cite{fukyan}, further developed in
\cite{leptoother}. We report here 
the results of Buchm\"uller  and  Pl\"umacher\cite{bupl}. 
Consider the generic leptonic Yukawa couplings
\be
{\cal L}_{Yuk}\ =\ -{\bar l_L}{\tilde \phi}g_l e_R 
		 -{\bar l_L}\phi g_{\nu} {\nu}_R
                 - \frac{1}{2} {\bar {\nu_R^c}}M {\nu}_R + h.c.
\label{eq:yuk}
\ee

This results in mass $m_l=g_lv$ for the charged lepton and
Dirac mass $m_D=g_{\nu}v$ for the neutrino if $\vev{\phi}=v$.
$M$ results from unknown physics but which is a \CP\  violating 
coupling. By see-saw mechanism we get $m_\nu\sim m_D^2/M$ and 
for the heavy species, $m_N\sim M$. 

Consider the decays of the
$N$ as in fig. \ref{fig:nudecay}, taken from \cite{bupl}.
From the interference of these diagrams (see also \cite{utpal}) 
we get for  the quantity $\cal B$ of eqn. (\ref{eq:bformula}),
\be
{\cal B}\ =\ {r - {\bar r} \over {r+ {\bar r}}}\  \sim\  {1\over v^2 m_D^2} {\rm Im} (m_D^{\dag}
m_D)^2
\label{eq:lasymm}
\ee

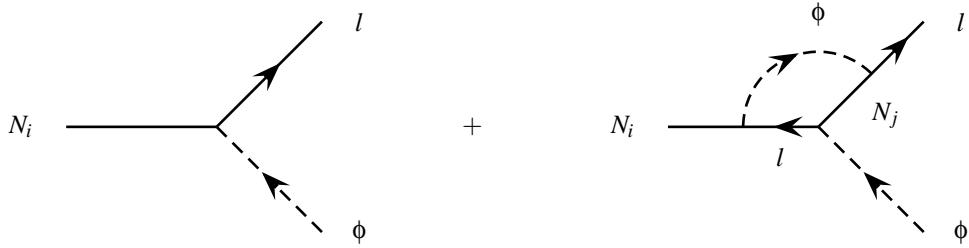
\begin{figure}
\input{Fig02.tex}
\caption{The contributions to $\nu$ decay whose interference 
results in $CP$ violating rates}
\label{fig:nudecay}
\end{figure}
In a model with $M\sim10^{14}$ GeV and with the assumption that
the neutrinos obey the hierarchy $m_{\nu_\mu}\sim10^{-3}$eV,
 $m_{\nu_e}\sim10^{-5}$eV  and $m_{\nu_\tau}\sim 0.1$eV,
it has been shown that $\Omega\sim 10^{-10}$ is possible to 
achieve. 
\end{description}

\section{Conclusion}\label{sec:conclusion}
All the physics needed for arriving at the important phenomenon of
sphalerons and of the high temperature anomaly was known since
mid-70's.  However a systematic investigation did not begin
until the mid-80's. The intensive development of many conceptual
issues and of calculation techniques now puts us in a position
to veto models of Particle Physics based on their potential for
dynamical explanation of the Baryon asymmetry. Some of the important 
techniques include improved calculations of the effective potential 
for the study of the nature of phase transition and the evolution 
of true vacuum bubbles as sites of \B-symmetry generation. 
These methods however are perturbative in nature and
if the problems are really hard lattice methods may be inevitable.

The generic but non-supersymmetric 2HDM is adequate for
\B-genesis at the electroweak scale. This however is not
natural in any unification scheme. The MSSM has been explored
extensively and this has resulted in constraints on
$\tan\beta$, and masses of the lightest Higgs and the stop.
The other very promising alternative is Leptogenesis, since
it may help to constrain neutrino physics, for which
interesting experimental evidence is emerging.

A generic mechanism applicable to SUSY GUTs and Supergravity
models is the Affleck-Dine mechanism\cite{affdin}. It bears investigation
in specific models, where it may place specific constraints on
the model. It was not possible to discuss this here
due to lack of space. 

Finally, the reader is referred to some of the more extensive 
recent reviews such as \cite{troddenrev}\cite{rubsha}\cite{riottorev}.

\section*{Acknowledgment}
The author wishes to thank the organisers of the DAE Symposium
at the Punjab University, Chandigarh. This work is
supported by a grant from the Department of Science and Technology.

\end{document}

%% file: vacgge.tex
\setlength{\unitlength}{0.240900pt}
\ifx\plotpoint\undefined\newsavebox{\plotpoint}\fi
\sbox{\plotpoint}{\rule[-0.200pt]{0.400pt}{0.400pt}}%
\begin{picture}(1500,749)(0,0)
\font\gnuplot=cmr10 at 10pt
\gnuplot
\sbox{\plotpoint}{\rule[-0.200pt]{0.400pt}{0.400pt}}%
\put(220.0,113.0){\rule[-0.200pt]{292.934pt}{0.400pt}}
\put(706.0,113.0){\rule[-0.200pt]{0.400pt}{136.831pt}}
\put(242.0,113.0){\rule[-0.200pt]{0.400pt}{4.818pt}}
\put(242,68){\makebox(0,0){-1.5}}
\put(242.0,661.0){\rule[-0.200pt]{0.400pt}{4.818pt}}
\put(397.0,113.0){\rule[-0.200pt]{0.400pt}{4.818pt}}
\put(397,68){\makebox(0,0){-1}}
\put(397.0,661.0){\rule[-0.200pt]{0.400pt}{4.818pt}}
\put(552.0,113.0){\rule[-0.200pt]{0.400pt}{4.818pt}}
\put(552,68){\makebox(0,0){-0.5}}
\put(552.0,661.0){\rule[-0.200pt]{0.400pt}{4.818pt}}
\put(706.0,113.0){\rule[-0.200pt]{0.400pt}{4.818pt}}
\put(706,68){\makebox(0,0){0}}
\put(706.0,661.0){\rule[-0.200pt]{0.400pt}{4.818pt}}
\put(861.0,113.0){\rule[-0.200pt]{0.400pt}{4.818pt}}
\put(861,68){\makebox(0,0){0.5}}
\put(861.0,661.0){\rule[-0.200pt]{0.400pt}{4.818pt}}
\put(1016.0,113.0){\rule[-0.200pt]{0.400pt}{4.818pt}}
\put(1016,68){\makebox(0,0){1}}
\put(1016.0,661.0){\rule[-0.200pt]{0.400pt}{4.818pt}}
\put(1171.0,113.0){\rule[-0.200pt]{0.400pt}{4.818pt}}
\put(1171,68){\makebox(0,0){1.5}}
\put(1171.0,661.0){\rule[-0.200pt]{0.400pt}{4.818pt}}
\put(1326.0,113.0){\rule[-0.200pt]{0.400pt}{4.818pt}}
\put(1326,68){\makebox(0,0){2}}
\put(1326.0,661.0){\rule[-0.200pt]{0.400pt}{4.818pt}}
\put(45,397){\makebox(0,0){$E$}}
\put(828,23){\makebox(0,0){$N_{C-S}[A^{\mu}(x)]$}}
\put(828,726){\makebox(0,0){ }}
\put(220,529){\usebox{\plotpoint}}
\multiput(220.58,529.00)(0.492,0.712){21}{\rule{0.119pt}{0.667pt}}
\multiput(219.17,529.00)(12.000,15.616){2}{\rule{0.400pt}{0.333pt}}
\multiput(232.00,546.60)(1.797,0.468){5}{\rule{1.400pt}{0.113pt}}
\multiput(232.00,545.17)(10.094,4.000){2}{\rule{0.700pt}{0.400pt}}
\multiput(245.00,548.92)(0.600,-0.491){17}{\rule{0.580pt}{0.118pt}}
\multiput(245.00,549.17)(10.796,-10.000){2}{\rule{0.290pt}{0.400pt}}
\multiput(257.58,536.40)(0.492,-0.970){21}{\rule{0.119pt}{0.867pt}}
\multiput(256.17,538.20)(12.000,-21.201){2}{\rule{0.400pt}{0.433pt}}
\multiput(269.58,511.88)(0.492,-1.444){21}{\rule{0.119pt}{1.233pt}}
\multiput(268.17,514.44)(12.000,-31.440){2}{\rule{0.400pt}{0.617pt}}
\multiput(281.58,477.09)(0.493,-1.686){23}{\rule{0.119pt}{1.423pt}}
\multiput(280.17,480.05)(13.000,-40.046){2}{\rule{0.400pt}{0.712pt}}
\multiput(294.58,432.67)(0.492,-2.133){21}{\rule{0.119pt}{1.767pt}}
\multiput(293.17,436.33)(12.000,-46.333){2}{\rule{0.400pt}{0.883pt}}
\multiput(306.58,382.11)(0.492,-2.306){21}{\rule{0.119pt}{1.900pt}}
\multiput(305.17,386.06)(12.000,-50.056){2}{\rule{0.400pt}{0.950pt}}
\multiput(318.58,328.69)(0.493,-2.122){23}{\rule{0.119pt}{1.762pt}}
\multiput(317.17,332.34)(13.000,-50.344){2}{\rule{0.400pt}{0.881pt}}
\multiput(331.58,274.53)(0.492,-2.176){21}{\rule{0.119pt}{1.800pt}}
\multiput(330.17,278.26)(12.000,-47.264){2}{\rule{0.400pt}{0.900pt}}
\multiput(343.58,224.36)(0.492,-1.918){21}{\rule{0.119pt}{1.600pt}}
\multiput(342.17,227.68)(12.000,-41.679){2}{\rule{0.400pt}{0.800pt}}
\multiput(355.58,180.74)(0.492,-1.487){21}{\rule{0.119pt}{1.267pt}}
\multiput(354.17,183.37)(12.000,-32.371){2}{\rule{0.400pt}{0.633pt}}
\multiput(367.58,147.39)(0.493,-0.972){23}{\rule{0.119pt}{0.869pt}}
\multiput(366.17,149.20)(13.000,-23.196){2}{\rule{0.400pt}{0.435pt}}
\multiput(380.00,124.92)(0.496,-0.492){21}{\rule{0.500pt}{0.119pt}}
\multiput(380.00,125.17)(10.962,-12.000){2}{\rule{0.250pt}{0.400pt}}
\put(392,114.17){\rule{2.500pt}{0.400pt}}
\multiput(392.00,113.17)(6.811,2.000){2}{\rule{1.250pt}{0.400pt}}
\multiput(404.58,116.00)(0.493,0.536){23}{\rule{0.119pt}{0.531pt}}
\multiput(403.17,116.00)(13.000,12.898){2}{\rule{0.400pt}{0.265pt}}
\multiput(417.58,130.00)(0.492,1.186){21}{\rule{0.119pt}{1.033pt}}
\multiput(416.17,130.00)(12.000,25.855){2}{\rule{0.400pt}{0.517pt}}
\multiput(429.58,158.00)(0.492,1.616){21}{\rule{0.119pt}{1.367pt}}
\multiput(428.17,158.00)(12.000,35.163){2}{\rule{0.400pt}{0.683pt}}
\multiput(441.58,196.00)(0.492,1.961){21}{\rule{0.119pt}{1.633pt}}
\multiput(440.17,196.00)(12.000,42.610){2}{\rule{0.400pt}{0.817pt}}
\multiput(453.58,242.00)(0.493,2.043){23}{\rule{0.119pt}{1.700pt}}
\multiput(452.17,242.00)(13.000,48.472){2}{\rule{0.400pt}{0.850pt}}
\multiput(466.58,294.00)(0.492,2.306){21}{\rule{0.119pt}{1.900pt}}
\multiput(465.17,294.00)(12.000,50.056){2}{\rule{0.400pt}{0.950pt}}
\multiput(478.58,348.00)(0.492,2.263){21}{\rule{0.119pt}{1.867pt}}
\multiput(477.17,348.00)(12.000,49.126){2}{\rule{0.400pt}{0.933pt}}
\multiput(490.58,401.00)(0.493,1.924){23}{\rule{0.119pt}{1.608pt}}
\multiput(489.17,401.00)(13.000,45.663){2}{\rule{0.400pt}{0.804pt}}
\multiput(503.58,450.00)(0.492,1.789){21}{\rule{0.119pt}{1.500pt}}
\multiput(502.17,450.00)(12.000,38.887){2}{\rule{0.400pt}{0.750pt}}
\multiput(515.58,492.00)(0.492,1.315){21}{\rule{0.119pt}{1.133pt}}
\multiput(514.17,492.00)(12.000,28.648){2}{\rule{0.400pt}{0.567pt}}
\multiput(527.58,523.00)(0.492,0.841){21}{\rule{0.119pt}{0.767pt}}
\multiput(526.17,523.00)(12.000,18.409){2}{\rule{0.400pt}{0.383pt}}
\multiput(539.00,543.59)(0.950,0.485){11}{\rule{0.843pt}{0.117pt}}
\multiput(539.00,542.17)(11.251,7.000){2}{\rule{0.421pt}{0.400pt}}
\multiput(552.00,548.93)(0.874,-0.485){11}{\rule{0.786pt}{0.117pt}}
\multiput(552.00,549.17)(10.369,-7.000){2}{\rule{0.393pt}{0.400pt}}
\multiput(564.58,539.82)(0.492,-0.841){21}{\rule{0.119pt}{0.767pt}}
\multiput(563.17,541.41)(12.000,-18.409){2}{\rule{0.400pt}{0.383pt}}
\multiput(576.58,518.16)(0.492,-1.358){21}{\rule{0.119pt}{1.167pt}}
\multiput(575.17,520.58)(12.000,-29.579){2}{\rule{0.400pt}{0.583pt}}
\multiput(588.58,485.35)(0.493,-1.607){23}{\rule{0.119pt}{1.362pt}}
\multiput(587.17,488.17)(13.000,-38.174){2}{\rule{0.400pt}{0.681pt}}
\multiput(601.58,442.80)(0.492,-2.090){21}{\rule{0.119pt}{1.733pt}}
\multiput(600.17,446.40)(12.000,-45.402){2}{\rule{0.400pt}{0.867pt}}
\multiput(613.58,393.25)(0.492,-2.263){21}{\rule{0.119pt}{1.867pt}}
\multiput(612.17,397.13)(12.000,-49.126){2}{\rule{0.400pt}{0.933pt}}
\multiput(625.58,340.56)(0.493,-2.162){23}{\rule{0.119pt}{1.792pt}}
\multiput(624.17,344.28)(13.000,-51.280){2}{\rule{0.400pt}{0.896pt}}
\multiput(638.58,285.53)(0.492,-2.176){21}{\rule{0.119pt}{1.800pt}}
\multiput(637.17,289.26)(12.000,-47.264){2}{\rule{0.400pt}{0.900pt}}
\multiput(650.58,235.08)(0.492,-2.004){21}{\rule{0.119pt}{1.667pt}}
\multiput(649.17,238.54)(12.000,-43.541){2}{\rule{0.400pt}{0.833pt}}
\multiput(662.58,189.33)(0.492,-1.616){21}{\rule{0.119pt}{1.367pt}}
\multiput(661.17,192.16)(12.000,-35.163){2}{\rule{0.400pt}{0.683pt}}
\multiput(674.58,153.14)(0.493,-1.052){23}{\rule{0.119pt}{0.931pt}}
\multiput(673.17,155.07)(13.000,-25.068){2}{\rule{0.400pt}{0.465pt}}
\multiput(687.58,127.51)(0.492,-0.625){21}{\rule{0.119pt}{0.600pt}}
\multiput(686.17,128.75)(12.000,-13.755){2}{\rule{0.400pt}{0.300pt}}
\put(699,113.67){\rule{2.891pt}{0.400pt}}
\multiput(699.00,114.17)(6.000,-1.000){2}{\rule{1.445pt}{0.400pt}}
\multiput(711.00,114.58)(0.539,0.492){21}{\rule{0.533pt}{0.119pt}}
\multiput(711.00,113.17)(11.893,12.000){2}{\rule{0.267pt}{0.400pt}}
\multiput(724.58,126.00)(0.492,1.056){21}{\rule{0.119pt}{0.933pt}}
\multiput(723.17,126.00)(12.000,23.063){2}{\rule{0.400pt}{0.467pt}}
\multiput(736.58,151.00)(0.492,1.530){21}{\rule{0.119pt}{1.300pt}}
\multiput(735.17,151.00)(12.000,33.302){2}{\rule{0.400pt}{0.650pt}}
\multiput(748.58,187.00)(0.492,1.918){21}{\rule{0.119pt}{1.600pt}}
\multiput(747.17,187.00)(12.000,41.679){2}{\rule{0.400pt}{0.800pt}}
\multiput(760.58,232.00)(0.493,2.003){23}{\rule{0.119pt}{1.669pt}}
\multiput(759.17,232.00)(13.000,47.535){2}{\rule{0.400pt}{0.835pt}}
\multiput(773.58,283.00)(0.492,2.306){21}{\rule{0.119pt}{1.900pt}}
\multiput(772.17,283.00)(12.000,50.056){2}{\rule{0.400pt}{0.950pt}}
\multiput(785.58,337.00)(0.492,2.263){21}{\rule{0.119pt}{1.867pt}}
\multiput(784.17,337.00)(12.000,49.126){2}{\rule{0.400pt}{0.933pt}}
\multiput(797.58,390.00)(0.493,2.003){23}{\rule{0.119pt}{1.669pt}}
\multiput(796.17,390.00)(13.000,47.535){2}{\rule{0.400pt}{0.835pt}}
\multiput(810.58,441.00)(0.492,1.832){21}{\rule{0.119pt}{1.533pt}}
\multiput(809.17,441.00)(12.000,39.817){2}{\rule{0.400pt}{0.767pt}}
\multiput(822.58,484.00)(0.492,1.444){21}{\rule{0.119pt}{1.233pt}}
\multiput(821.17,484.00)(12.000,31.440){2}{\rule{0.400pt}{0.617pt}}
\multiput(834.58,518.00)(0.492,0.927){21}{\rule{0.119pt}{0.833pt}}
\multiput(833.17,518.00)(12.000,20.270){2}{\rule{0.400pt}{0.417pt}}
\multiput(846.00,540.58)(0.652,0.491){17}{\rule{0.620pt}{0.118pt}}
\multiput(846.00,539.17)(11.713,10.000){2}{\rule{0.310pt}{0.400pt}}
\multiput(859.00,548.94)(1.651,-0.468){5}{\rule{1.300pt}{0.113pt}}
\multiput(859.00,549.17)(9.302,-4.000){2}{\rule{0.650pt}{0.400pt}}
\multiput(871.58,543.09)(0.492,-0.755){21}{\rule{0.119pt}{0.700pt}}
\multiput(870.17,544.55)(12.000,-16.547){2}{\rule{0.400pt}{0.350pt}}
\multiput(883.58,523.88)(0.493,-1.131){23}{\rule{0.119pt}{0.992pt}}
\multiput(882.17,525.94)(13.000,-26.940){2}{\rule{0.400pt}{0.496pt}}
\multiput(896.58,493.05)(0.492,-1.703){21}{\rule{0.119pt}{1.433pt}}
\multiput(895.17,496.03)(12.000,-37.025){2}{\rule{0.400pt}{0.717pt}}
\multiput(908.58,452.08)(0.492,-2.004){21}{\rule{0.119pt}{1.667pt}}
\multiput(907.17,455.54)(12.000,-43.541){2}{\rule{0.400pt}{0.833pt}}
\multiput(920.58,404.25)(0.492,-2.263){21}{\rule{0.119pt}{1.867pt}}
\multiput(919.17,408.13)(12.000,-49.126){2}{\rule{0.400pt}{0.933pt}}
\multiput(932.58,351.69)(0.493,-2.122){23}{\rule{0.119pt}{1.762pt}}
\multiput(931.17,355.34)(13.000,-50.344){2}{\rule{0.400pt}{0.881pt}}
\multiput(945.58,297.25)(0.492,-2.263){21}{\rule{0.119pt}{1.867pt}}
\multiput(944.17,301.13)(12.000,-49.126){2}{\rule{0.400pt}{0.933pt}}
\multiput(957.58,244.94)(0.492,-2.047){21}{\rule{0.119pt}{1.700pt}}
\multiput(956.17,248.47)(12.000,-44.472){2}{\rule{0.400pt}{0.850pt}}
\multiput(969.58,198.48)(0.493,-1.567){23}{\rule{0.119pt}{1.331pt}}
\multiput(968.17,201.24)(13.000,-37.238){2}{\rule{0.400pt}{0.665pt}}
\multiput(982.58,159.57)(0.492,-1.229){21}{\rule{0.119pt}{1.067pt}}
\multiput(981.17,161.79)(12.000,-26.786){2}{\rule{0.400pt}{0.533pt}}
\multiput(994.58,132.09)(0.492,-0.755){21}{\rule{0.119pt}{0.700pt}}
\multiput(993.17,133.55)(12.000,-16.547){2}{\rule{0.400pt}{0.350pt}}
\multiput(1006.00,115.94)(1.651,-0.468){5}{\rule{1.300pt}{0.113pt}}
\multiput(1006.00,116.17)(9.302,-4.000){2}{\rule{0.650pt}{0.400pt}}
\multiput(1018.00,113.58)(0.652,0.491){17}{\rule{0.620pt}{0.118pt}}
\multiput(1018.00,112.17)(11.713,10.000){2}{\rule{0.310pt}{0.400pt}}
\multiput(1031.58,123.00)(0.492,0.927){21}{\rule{0.119pt}{0.833pt}}
\multiput(1030.17,123.00)(12.000,20.270){2}{\rule{0.400pt}{0.417pt}}
\multiput(1043.58,145.00)(0.492,1.401){21}{\rule{0.119pt}{1.200pt}}
\multiput(1042.17,145.00)(12.000,30.509){2}{\rule{0.400pt}{0.600pt}}
\multiput(1055.58,178.00)(0.493,1.726){23}{\rule{0.119pt}{1.454pt}}
\multiput(1054.17,178.00)(13.000,40.982){2}{\rule{0.400pt}{0.727pt}}
\multiput(1068.58,222.00)(0.492,2.133){21}{\rule{0.119pt}{1.767pt}}
\multiput(1067.17,222.00)(12.000,46.333){2}{\rule{0.400pt}{0.883pt}}
\multiput(1080.58,272.00)(0.492,2.263){21}{\rule{0.119pt}{1.867pt}}
\multiput(1079.17,272.00)(12.000,49.126){2}{\rule{0.400pt}{0.933pt}}
\multiput(1092.58,325.00)(0.492,2.306){21}{\rule{0.119pt}{1.900pt}}
\multiput(1091.17,325.00)(12.000,50.056){2}{\rule{0.400pt}{0.950pt}}
\multiput(1104.58,379.00)(0.493,2.003){23}{\rule{0.119pt}{1.669pt}}
\multiput(1103.17,379.00)(13.000,47.535){2}{\rule{0.400pt}{0.835pt}}
\multiput(1117.58,430.00)(0.492,1.918){21}{\rule{0.119pt}{1.600pt}}
\multiput(1116.17,430.00)(12.000,41.679){2}{\rule{0.400pt}{0.800pt}}
\multiput(1129.58,475.00)(0.492,1.573){21}{\rule{0.119pt}{1.333pt}}
\multiput(1128.17,475.00)(12.000,34.233){2}{\rule{0.400pt}{0.667pt}}
\multiput(1141.58,512.00)(0.492,1.013){21}{\rule{0.119pt}{0.900pt}}
\multiput(1140.17,512.00)(12.000,22.132){2}{\rule{0.400pt}{0.450pt}}
\multiput(1153.00,536.58)(0.497,0.493){23}{\rule{0.500pt}{0.119pt}}
\multiput(1153.00,535.17)(11.962,13.000){2}{\rule{0.250pt}{0.400pt}}
\put(1166,547.67){\rule{2.891pt}{0.400pt}}
\multiput(1166.00,548.17)(6.000,-1.000){2}{\rule{1.445pt}{0.400pt}}
\multiput(1178.58,545.51)(0.492,-0.625){21}{\rule{0.119pt}{0.600pt}}
\multiput(1177.17,546.75)(12.000,-13.755){2}{\rule{0.400pt}{0.300pt}}
\multiput(1190.58,529.14)(0.493,-1.052){23}{\rule{0.119pt}{0.931pt}}
\multiput(1189.17,531.07)(13.000,-25.068){2}{\rule{0.400pt}{0.465pt}}
\multiput(1203.58,500.33)(0.492,-1.616){21}{\rule{0.119pt}{1.367pt}}
\multiput(1202.17,503.16)(12.000,-35.163){2}{\rule{0.400pt}{0.683pt}}
\multiput(1215.58,461.22)(0.492,-1.961){21}{\rule{0.119pt}{1.633pt}}
\multiput(1214.17,464.61)(12.000,-42.610){2}{\rule{0.400pt}{0.817pt}}
\multiput(1227.58,414.39)(0.492,-2.219){21}{\rule{0.119pt}{1.833pt}}
\multiput(1226.17,418.19)(12.000,-48.195){2}{\rule{0.400pt}{0.917pt}}
\multiput(1239.58,362.69)(0.493,-2.122){23}{\rule{0.119pt}{1.762pt}}
\multiput(1238.17,366.34)(13.000,-50.344){2}{\rule{0.400pt}{0.881pt}}
\multiput(1252.58,308.25)(0.492,-2.263){21}{\rule{0.119pt}{1.867pt}}
\multiput(1251.17,312.13)(12.000,-49.126){2}{\rule{0.400pt}{0.933pt}}
\multiput(1264.58,255.80)(0.492,-2.090){21}{\rule{0.119pt}{1.733pt}}
\multiput(1263.17,259.40)(12.000,-45.402){2}{\rule{0.400pt}{0.867pt}}
\multiput(1276.58,208.22)(0.493,-1.646){23}{\rule{0.119pt}{1.392pt}}
\multiput(1275.17,211.11)(13.000,-39.110){2}{\rule{0.400pt}{0.696pt}}
\multiput(1289.58,167.16)(0.492,-1.358){21}{\rule{0.119pt}{1.167pt}}
\multiput(1288.17,169.58)(12.000,-29.579){2}{\rule{0.400pt}{0.583pt}}
\multiput(1301.58,136.82)(0.492,-0.841){21}{\rule{0.119pt}{0.767pt}}
\multiput(1300.17,138.41)(12.000,-18.409){2}{\rule{0.400pt}{0.383pt}}
\multiput(1313.00,118.93)(0.874,-0.485){11}{\rule{0.786pt}{0.117pt}}
\multiput(1313.00,119.17)(10.369,-7.000){2}{\rule{0.393pt}{0.400pt}}
\multiput(1325.00,113.59)(1.123,0.482){9}{\rule{0.967pt}{0.116pt}}
\multiput(1325.00,112.17)(10.994,6.000){2}{\rule{0.483pt}{0.400pt}}
\multiput(1338.58,119.00)(0.492,0.841){21}{\rule{0.119pt}{0.767pt}}
\multiput(1337.17,119.00)(12.000,18.409){2}{\rule{0.400pt}{0.383pt}}
\multiput(1350.58,139.00)(0.492,1.358){21}{\rule{0.119pt}{1.167pt}}
\multiput(1349.17,139.00)(12.000,29.579){2}{\rule{0.400pt}{0.583pt}}
\multiput(1362.58,171.00)(0.493,1.607){23}{\rule{0.119pt}{1.362pt}}
\multiput(1361.17,171.00)(13.000,38.174){2}{\rule{0.400pt}{0.681pt}}
\multiput(1375.58,212.00)(0.492,2.090){21}{\rule{0.119pt}{1.733pt}}
\multiput(1374.17,212.00)(12.000,45.402){2}{\rule{0.400pt}{0.867pt}}
\multiput(1387.58,261.00)(0.492,2.263){21}{\rule{0.119pt}{1.867pt}}
\multiput(1386.17,261.00)(12.000,49.126){2}{\rule{0.400pt}{0.933pt}}
\multiput(1399.58,314.00)(0.492,2.306){21}{\rule{0.119pt}{1.900pt}}
\multiput(1398.17,314.00)(12.000,50.056){2}{\rule{0.400pt}{0.950pt}}
\multiput(1411.58,368.00)(0.493,2.043){23}{\rule{0.119pt}{1.700pt}}
\multiput(1410.17,368.00)(13.000,48.472){2}{\rule{0.400pt}{0.850pt}}
\multiput(1424.58,420.00)(0.492,2.004){21}{\rule{0.119pt}{1.667pt}}
\multiput(1423.17,420.00)(12.000,43.541){2}{\rule{0.400pt}{0.833pt}}
\end{picture}

%% file: Fig02.tex
\begin{center}
  \pspicture(1,0)(13.5,4)
    \psline[linewidth=1pt](1.6,2)(3.6,2)
    \psline[linewidth=1pt](3.6,2)(5,3.4)
    \psline[linewidth=1pt,linestyle=dashed](3.6,2)(5,0.6)
    \psline[linewidth=2pt]{<-}(4.45,2.85)(4.25,2.65)
    \psline[linewidth=2pt]{->}(4.4,1.2)(4.2,1.4)
    \rput[cc]{0}(1,2){$\displaystyle N_i$}
    \rput[cc]{0}(5.5,3.4){$\displaystyle l$}
    \rput[cc]{0}(5.5,0.6){$\displaystyle \phi$}
    \rput[cc]{0}(7,2){$+$}
    \psline[linewidth=1pt](9.6,2)(11.6,2)
    \psline[linewidth=1pt](11.6,2)(13,3.4)
    \psline[linewidth=1pt,linestyle=dashed](11.6,2)(13,0.6)
    \psarc[linewidth=1pt,linestyle=dashed](11.6,2){1cm}{45}{180}
    \psline[linewidth=2pt]{<-}(12.80,3.20)(12.60,3.00)
    \psline[linewidth=2pt]{->}(12.4,1.2)(12.2,1.4)
    \psline[linewidth=2pt]{<-}(11.0,2.0)(11.2,2.0)
    \psline[linewidth=2pt]{->}(11.14,2.88)(11.34,2.98)
    \rput[cc]{0}(9,2){$\displaystyle N_i$}
    \rput[cc]{0}(13.5,3.4){$\displaystyle l$}
    \rput[cc]{0}(13.5,0.6){$\displaystyle \phi$}
    \rput[cc]{0}(11.6,3.5){$\phi$}
    \rput[cc]{0}(12.5,2.15){$N_j$}
    \rput[cc]{0}(11.1,1.6){$l$}
  \endpspicture
\end{center}

%% file: dbg.bbl
\begin{thebibliography}{99}


\bibitem{wein1}
S. Weinberg, in ``{\sl Lectures on Particles and Fields}'', edited by
K. Johnson at al, (Prentice-Hall, Englewood Cliffs, N.J., 1964),
pg. 482.

\bibitem{sakh} A. D. Sakharov, {\sl JETP Lett.} {\bf 5}, 24 (1967)

\bibitem{yosh} M. Yoshimura, \PRL{41}, 281 (1978); {\it erratum}, 
{\bf 42}, 740 (1979)  

\bibitem{wein2} S. Weinberg, \PRL{42}, 850 (1979)

\bibitem{fukyan} M. Fukugita and T. Yanagida \PLB{174}, 45 (1986);
\PRD{42} 1285 (1990)



\bibitem{vsoni} V. Soni, \PLB{93} 101 (1980)

\bibitem{kliman} F. R. Klinkhammer and N. S. Manton, \PRD{30}, 2212 (1984)

\bibitem{brihaye} Y. Brihaye, B. Kleihaus and J. Kunz \PRD{46} 3587 (1992); 
\PRD{47} 1664 (1993); S. Braibant, Y. Brihaye and J. Kunz
{\sl Int. J. Mod. Phy.} {\bf A8} 5563 (1993) 

\bibitem{armc}  
P. Arnold  and McLerran L., \PRD{36}, 581 (1987); \PRD{37}, 1020 (1988)

\bibitem{mhbound}  Bochkarev A. I. and Shaposhnikov M. E., 
{\sl Mod. Phys. Lett.} {\bf A2}, 417 (1987); 
Bochkarev A. I., Khlebnikov S. Yu.
and Shaposhnikov M.E., \NPB{329}, 490 (1990)

\bibitem{moorelat} G. D. Moore \PRD{59} 014503 (1999); see also
{\it hep-lat/9808031}.

\bibitem{aaps}
J. Ambjorn, Askgaard t., Porter H. and Shaposhnikov M.
E., \PLB{244},  479 (1990); \NPB{353}, 346 (1991)

\bibitem{krs} V. A. Kuzmin, V. A. Rubakov, and M. E. Shaposhnikov,
\PLB{155}, 36 (1985)

\bibitem{colbub}  S. Coleman, \PRD{15}, 2929 (1977); C. Callan and
S. Coleman \PRD{16}, 1762 (1977).


\bibitem{linbub} Linde A. D., \PLB{70}, 306 (1977); \PLB{100}, 37 (1981);
\NPB{216}, 421 (1983)

\bibitem{jarlskog} C. Jarlskog, \PRL{55}, 1039 (1985)

\bibitem{arnold} P. Arnold in {\sl ``Electroweak Physics and the
Early Universe", J. C. Romao and F. Freire, ed.s, NATO ASI series
{\bf B} vol. 338, Plenum Press, 1994, pp. 79-91 }

\bibitem{arn2loop} P. Arnold and J. R. Espinosa \PRD{47}, 3546 (1993)

\bibitem{buch2loop} W. Buchm\"uller, Z. Fodor and A. Hebecker \NPB{447}, 
317  (1995); See also W. Buch\"uller, Z. Fodor, T. Helbig and
D. Walliser {\sl Ann. Phys.} {\bf 260} (1994).

\bibitem{appisar} T. Applequist and R. Pisarski \PRD{23} 2305 (1981)

\bibitem{nadkarni} S. Nadkarni \PRD{27} 388 (1983)

\bibitem{faracetal} K. Faracos, K. Kajantie, K. Rummukainen
and M. Shaposhnikov \NPB{425}, 67 (1994); 
K. Kajantie, M. Laine, K. Rummukainen and M. Shaposhnikov, 
\NPB{458} 90 (1996)

\bibitem{andhal} G. E. Anderson and L. J. Hall, \PRD{45}, 2685 (1992)


\bibitem{cokaneprop} A. G. Cohen and D. B. Kaplan, \PLB{199}, 251 (1987);
{\NPB 308}, 913 (1988); A. G. Cohen, D. B. Kaplan and A. E. Nelson
{\PLB 263}, 86 (1991)

\bibitem{mstv} L. McLerran, M. Shaposhnikov, N. Turok and M.
Voloshin, {\PLB 256}, 351 (1991)

\bibitem{sbdanduay2} S. Bhowmik Duari and U. A. Yajnik, {\sl Nucl. 
Phys.} {\bf B} (proc. suppl.) {\bf 43} (1995) 282-285

\bibitem{cokanerev} A.G Cohen, D.B Kaplan and A.E Nelson {\it Ann. Rev. Nucl.
     Part. Sci.} {\bf 43},  27 (1993)

\bibitem{sbdanduay3} See for example S. B. Duari and U. A. Yajnik {\sl Mod.
Phy. Lett.} {\bf A11}, 2481 (1996) 

\bibitem{cqriw} M. Carena, M. Quiros, A. Riotto, I. Vilja
and C. E. M. Wagner \NPB{503} 387 (1997)

\bibitem{cqw} M. Carena, M. Quiros, C. E. M.  Wagner
\NPB{524} 3 (1998)

\bibitem{clijoka} J. Cline, M. Joyce and K. Kainulainen, \PLB{417} 
79 (1998) 

\bibitem{mooaqu} J. Moreno, D. Oaknin and M. Quiros
\NPB{483} 267 (1997)

\bibitem{sbdanduay1} S. Bhowmik Duari and U. A. Yajnik, \PLB 326 21 (1994) 

\bibitem{branden1} R. Brandenberger, A-C. Davis, and
M. Trodden, \PLB 335 123 (1994) 

\bibitem{troddenetal} M. Trodden, A-C. Daviis and R. H. Brandenberger
\PLB{349}  131, (1995)

\bibitem{branden96} R. H. Brandenberger, A-C. Davis, T. Prokopec 
and M. Trodden, \PRD{53} 4257 (1996)

\bibitem{mooreclineetal} J. Cline, J. Espinosa, G. D. Moore and A. Riotto,
\PRD{59} 069014 (1999)


\bibitem{mencoo} S. Ben-Menahem and A. S. Cooper, \NPB 388
(1992) 409

\bibitem{lewrio} H. Lew and A. Riotto, \PLB{309}, 258  (1993) 

\bibitem{ywmmc} U. A. Yajnik, H. Widyan, A. Mukherjee, S. Mahajan
and D. Choudhury 
\PRD{59}, 103508 (1999)\\
see also in {\sl ``Proceedings of WHEPP-5''}, R. Gavai and
R. Godbole, ed.s, {\sl Pramana} {\bf 51} 276 (1998) 


\bibitem{leptoother} P. langacker, R. Peccei and M. Yanagida
{\sl Mod. Phys. lett.} {\bf A1} 541 (1986)\\
M. Luty \PRD{45} 455 (1992)

\bibitem{bupl} W. Buchm\"uller and M. Pl\"umacher, \PLB{431}
354 (1998)

\bibitem{utpal} For a discussion of other possible contributions to
\CP\  violation see U. Sarkar {\it hep-ph/9810247}, contribution
to {\sl DARK98}, Heidelberg, July 1998 and references therein.

\bibitem{affdin} I. Affleck and M. Dine \NPB{249} 361 (1985)

\bibitem{troddenrev} M. Trodden {\sl Rev. Mod. Phys.} {\bf 71} 1463 (1999)

\bibitem{rubsha} V. A. Rubakov and M. E. Shaposhnikov, {\sl Phys. Uspekhi} 
{\bf 39} 461 (1996)

\bibitem{riottorev} A. Riotto, {\sl ``Theories of Baryogenesis''},  {\it
hep-ph/9807454}, Lectures at the
Summer School on HEP and Cosmology, ICTP, Trieste, 1998

\end{thebibliography}
